\documentclass{svproc}
\usepackage{url}
\usepackage{amsmath,amssymb}
\usepackage{multicol}
\usepackage{footmisc}
\usepackage{graphicx}
\usepackage{cite}

\begin{document}
\mainmatter              
\title{Success at high peaks: a multiscale approach combining individual and expedition-wide factors}
\titlerunning{Success at high peaks}  
%
\author{Sanjukta Krishnagopal}
\authorrunning{Sanjukta Krishnagopal}
\institute{Gatsby Computational Neuroscience Unit, University College London-W1T 4JG, UK.\\
\email{s.krishnagopal@ucl.ac.uk}} 
\maketitle              

\begin{abstract}
This work presents a network-based data-driven study of the combination of factors that contribute to success in mountaineering. It simultaneously examines the effects of individual factors such as age, gender, experience etc., as well as expedition-wide factors such as number of camps, ratio of sherpas to paying climbers etc. Specifically, it combines the two perspectives into a multiscale network, i.e., a network of individual climber features within each expedition at the finer scale, and an expedition similarity network on the coarser scale. The latter is represented as a multiplex network where layers encode different factors. 
The analysis reveals that chances of failure to summit due to fatigue, altitude or logistical problems, drastically reduce when climbing with repeat partners, especially for experienced climbers.
Additionally, node-centrality indicates that individual traits of youth and oxygen use are the strongest drivers of success. Further, the learning of network projections enables computation of correlations between intra-expedition networks and corresponding expedition success rates. Of expedition-wide factors, the expedition size and length layers are found to be strongly correlated with success rate. Lastly, community detection on the expedition-similarity network reveals distinct communities where a difference in success rates naturally emerges amongst the communities.

\end{abstract}
\section*{Introduction}

Extreme mountaineering is an increasingly popular activity that demands not only physical fitness and skills, but also mental fortitude and psychological control. Some of the impressive mountain ranges are Himalayas and present several opportunities, including the famous Mount Everest itself, for extreme mountaineering. 
Extreme or high altitude mountaineering is not what one might consider safe, and personal or expedition-related factors such as effective use of proper equipment, climber experience, mental strength and self-reliance are all measures \cite{schussman1990} to increase safety and chances of success. Certain aspects of extreme mountaineering are well-known to be individualistic, especially as one gets closer to the death zone (8000m altitude). However, with increasing commercialization of extreme mountaineering, social and psychological factors play a subtle but crucial role in survival. Indeed the mass fatality on Everest in 1996 received tremendous attention social and logistical misgivings of the expeditions \cite{tempest2007}. 

Understanding factors, both individual and expedition-wide e.g. effective use of proper equipment, climber experience, mental state etc. \cite{schussman1990}  is crucial to maximizing safety and chances of success. Data driven analysis of success-predicating factors has been accelerated by the availability of the large and detailed Himalayan dataset \cite{data04}. Indeed, several works have studied the effects of age and sex \cite{huey2007}, experience \cite{huey2020}, commercialization \cite{westhoff2012} etc. on success, and highlight the importance of age as a dominant determining factor. Additionally, \cite{weinbruch2013} shows that women are more risk-averse than men, and sherpas have lower risk at high altitude than paying climbers. 

Success depends both on both physiological state \cite{huey2001,szymczak2021}, as well as psychological and sociological state \cite{helms1984} that influence the evaluation of risks and hazards. In \cite{ewert1985}, the psychological motivators behind why people climb is outlined. These motivators differ for paying climbers vs sherpas introduces questions regarding the ethics of hiring sherpas \cite{nyaupane2015}. Group dynamics also play a major role in anxiety and problem solving \cite{tougne2008}. Thus, despite opinions that climbing is an individual activity, there is mounting evidence highlighting the importance of social and psychological factors. The psychology is largely driven by relationships between climbers, for instance climbers that frequently climb together may developer better group dynamics, and consequently lower failure. However, there is limited investigation of the effects of climbing with repeat partners. This work studies the likelihood of failing when climbing with repeat partners due to factors such as logistical failings, fatigue, altitude related sickness etc. 

Network science is becoming increasingly popular when studying data consisting of a multitude of complex interacting factors. Network approaches have been used successful in predictive medicine \cite{krishnagopal2021}, climate prediction \cite{steinhaeuser2011}, predictions in group sports \cite{lusher2010}, disease spreading \cite{disease2019} etc.
A mountaineering expedition consists of individuals which naturally lend themselves to a network structure. Relationships between individuals and expedition features such as oxygen use, age, sex, experience etc. can be modeled through a bipartite network. This network are often projected into other spaces for further analysis \cite{krishnagopal2020,larremore2014}. To the best of our knowledge, this work is the first network-based analysis of mountaineering data, incorporating factors at multiple scales.
The natural question emerges, which of these features, which can be represented by nodes, are central to maximizing chances of success? An active area of research investigates the importance of nodes \cite{mo2019} through centrality-based measures \cite{sola2013,saito2016} that serve as reliable indicators of `important' factors.

While several studies have focused on individual traits that affect success and death, it is natural to expect expedition-wide factors (e.g. ratio of sherpas to paying climbers, number of days to summit, and number of camps, intra-expedition social relationships etc.) to play a role in success. However there is limited work that studies the effect of such expedition-wide factors. This work considers both expedition-wide factors and personal features, and the interaction between the two. Multilayer networks \cite{kivela2014} are an ideal tool to model multiple types of interactions, where each layer models relationships between expeditions through a particular factor. Multilayer networks have been used successfully to model neuoronal activity \cite{vaiana2020}, in sports \cite{buldu2018}, in biomedicine \cite{hammoud2020} etc. 
Multilayer networks without intra-layer connections between different nodes are also known as multiplex networks \cite{lee2015}. In order to model the different factors that influence expedition similarities, a multiplex network model is used.
However the network of feature-relations within an expedition is in itself one of the factors correlated with success, hence one layer of the multilayer network encodes similarities in intra-expedition networks, where connectivity between expeditions is determined by their graph similarity \cite{gao2010}. This lends a multiscale structure to the network model. The term multiscale can be used to refer to different levels of thresholding in the graph across different scales as in \cite{pereira2019}, or hierarchical networks as in \cite{sarkar2013, krishnagopal2017}. The notion of multiscale used here is derived from the latter, where the nodes of the expedition similarity network are in fact networks themselves. Multiscale networks are natural when modeling relationships on different scales for instance in brain modeling \cite{betzel2017}, stock market \cite{pereira2019}, ecology \cite{lenormand2018} etc. 

Motivations, ability and psychologies vary amongst individuals, influencing people's perceptions \cite{pomfret2006} and strategies that may contribute to success. Hence, there must exist multiple strategies that consist of a different combination dominant factors, and a climber may be interested in the strategy that is best suited to them. Community detection \cite{newman2006} on the expedition similarity networks naturally partitions expeditions into groups that show high within-group similarity, where each group defines a strategy. Community detection is a useful tool in network analysis, and is an active area of research extended to multilayer networks \cite{huang2021}, multiscale networks \cite{sarkar2013} etc. This work identifies three groups, with one in particular correlated with high success rate, providing insight into the combination of factors that allow for safe and successful climbs.

\section{Data}

The data for this work was obtained from the open access Himalayan Database \cite{data04}, which is a compilation of records for all expeditions that have climbed in the Nepal Himalayan range. The dataset cover all expeditions from 1905 through 2021, and has records of 468 peaks, over 10,500 expedition records and over 78,400 records of climbers, where each record of any type is associated with an ID. 
We use the following information from the Expedition records:
\begin{itemize}
	\item Peak climbed (height).
	\item Days from basecamp to summit.
	\item Number of camps above basecamp.
	\item Total number of paying members and hired personnel.
	\item Result: (1) Success (main peak/foresummit/claimed), (2) No summit.
\end{itemize}
The success rate of an expedition is calculated as the fraction of members that succeeded. 
Each expedition comprises of several individual climbers yielding a natural multiscale structure. We use the following data about each climber:

\begin{itemize}
	\item Demographics: Age, Sex, Nationality.
	\item Oxygen use: ascending or descending.
	\item Previous experience above 8000m (calculated from all previous peak data for the climber).
	\item Result: 
	\begin{enumerate}
		\item Success
		\item Altitude related failure: Acute Mountain Sickness (AMS) symptoms, breathing problems, frostbite, snowblindness or coldness.
		\item Logistical or Planning failure: Lack of supplies, support or equipment problems, O2 system failure, too late in day or too slow, insufficient time left for expedition.
		\item Fatigue related failure: exhaustion, fatigue, weakness or lack of motivation.
		\item Accident related failure: death or injury to self or others.
	\end{enumerate}
\end{itemize}

\section{The effect of climbing with repeat partners}

\begin{figure}[htbp!]
	\centering
	\includegraphics[width=0.6\linewidth]{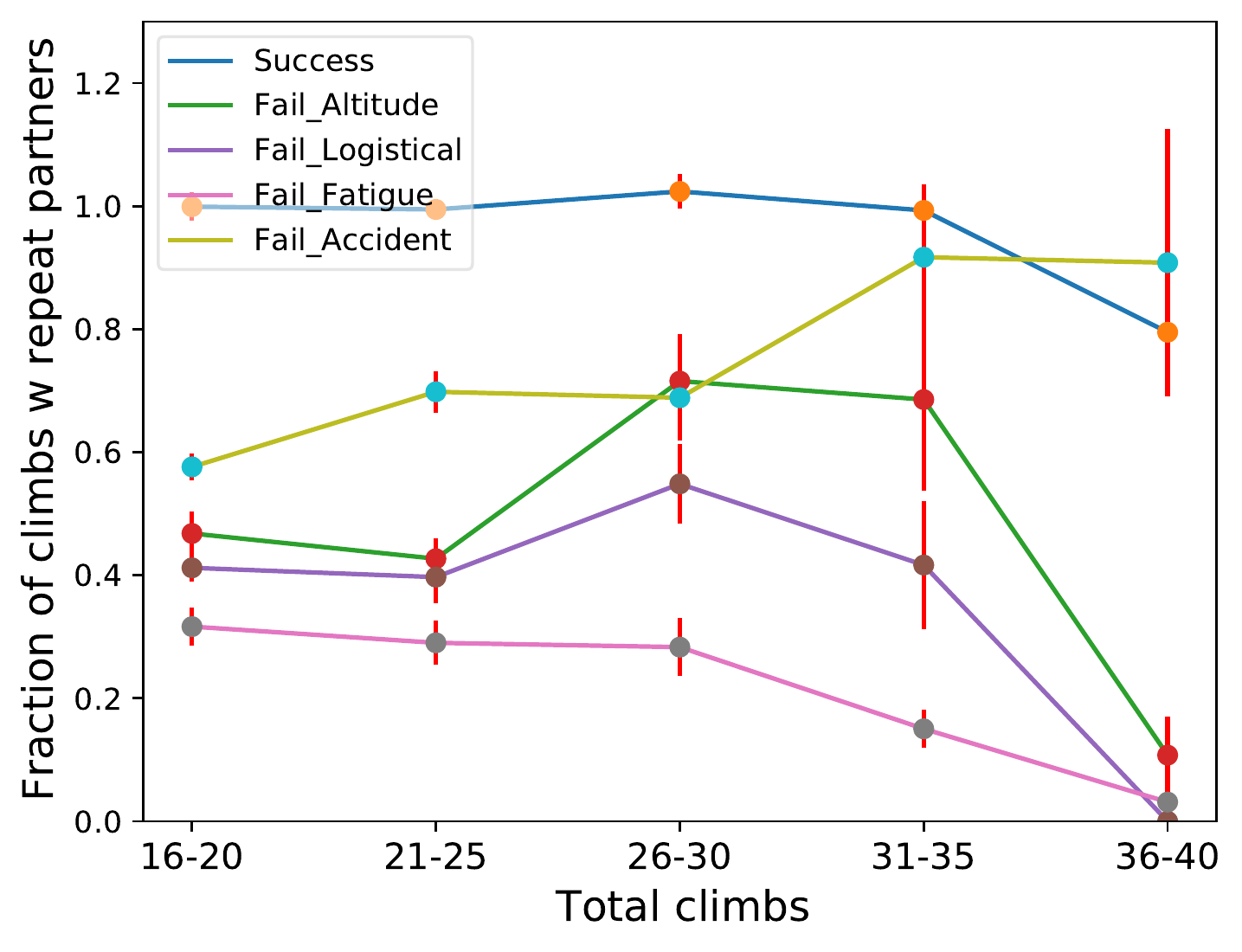}
	\caption{The fraction of success and various types of failures averaged over climbers when climbing with a group with at least one repeat partner (someone they have done a logged Himalaya expedition with before) over the individual average. The scores are presented as a function of the total number of climbs (listed within the dataset) of the individual.}
	\label{fig:friends}
\end{figure}

Climbers often tend to do repeat climbs with friends or regular climbing partners. The security of regular climbing partners may improve confidence and limit failure, but may also lead to a misleading sense of comfort. In Here, a comparison of average failure rate of a climber and failure when climbing with repeat partners is made. Fig.\ref{fig:friends} shows the fraction of failures when climbing with friends/repeat partners over the climber average. These failures are divided into altitude related, fatigue related, logistical and planning failures and accident/illness. The effect of total experience is normalized for by plotting across the total number of climbs on the x-axis starting at at least 15 climbs, hence not considering beginner climbers.

As seen in Fig.\ref{fig:friends}, repeat partners have virtually no effect on the chance of success except for very experienced climbers (36-40 logged climbs) which may be attributed to a increasing climb difficulty or of their partners being less experienced partners, both of which are more likely for very experienced climbers. In contrast, the chance of failure is significantly lower when climbing with repeat partners for every type of failure. In particular, the chance of failure due to fatigue-related issues is the most decreased when climbing with repeat partners, followed by failure due to logistical or planning issues. This may be expected since climbing partners that often climb together typically are better at communication, planing, and knowing each other's physical limitations. Note that only climbers with over 15 logged climbs are consider, indicating that complete lack of experience is not a cause of failure. Additionally, the most experienced climbers (that have logged 36-40 climbs) have nearly no failure due to fatigue or logistics, as one may expect. Similarly, failure due to altitude-related and cold-related issues also drastically reduces when climbing with repeat partners. Additionally, the cause of failure due to accident shows an increasing trend as a function of increasing experience, which may be attributed to the fact that more experienced climbers tend to tackle more dangerous mountains. 


\section{Intra-expedition features determining success}
\label{sec:centrality}
Here, the focus shifts from studying individual climbers to analyzing a group of climbers within an expedition. In order to do so, only the tallest peak, Mount Everest is considered. Expeditions with less than 12 climbers are excluded, as are expeditions that resulted in death. 
To generate the intra-expedition network, we start with a bipartite network $P$ between climbers and features, where a climber is connected to the features that they possess. The `features' selected as the nodes are: age, sex, oxygen while ascending, oxygen while descending, sherpa identity and previous experience about 8000m, making a total of $f=6$ features.
Each of this is a binary factor. A climber is connected to sex if they are male, and age is binarized into above and below median age $(40)$.

\begin{figure}[htbp!]
	\centering
	\includegraphics[width=0.95\linewidth]{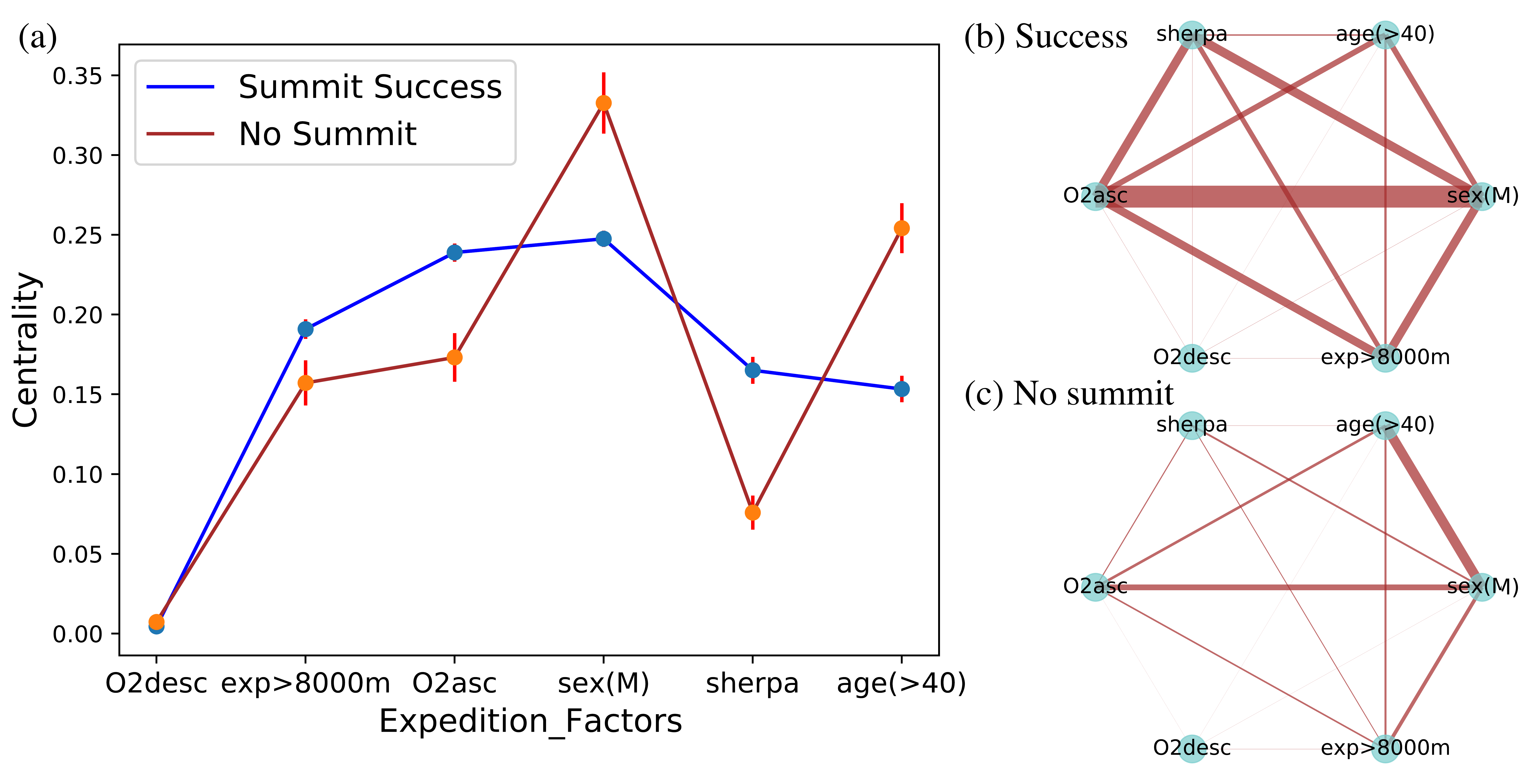}
	\caption{Mean eigenvector centrality (a) as a function of expedition features for Everest expeditions greater than 12 members plotted for groups of successful vs unsuccessful climbers ordered by increasing difference between success and no-success centralities. Error bars show standard error on the centrality. (b,c) Aggregation of the feature graph showing relative edge weights in summit success, and no-summit groups respectively.}
	\label{fig:centrality}
\end{figure}

We then generate the intra-expedition network $I$ of size $f \times f$ by projecting the bipartite network into feature space as follows: $I = P^T P$. The edge weight between two nodes (features) is given by the number of people that are connected to both the features.
Information about the expedition is then encoded in the structure of this network. To explore such effects, measures such as centrality capture important properties that provide insight into the importance of different features \cite{mo2019}. For instance, if the group were comprised of mostly high-age individuals, the node-centrality of the age node would be relatively high. Here, the eigenvector centrality \cite{sola2013} determines how central each feature is in a given graph.
Studying the differences in feature centrality between groups of successful summit vs no-summit provides important insight into features that may be important for summit success. 

As seen in Fig. \ref{fig:centrality} (a), the least central feature in determining success on summit was the use of oxygen while descending, which is expected since descent features have no effect on summit prospects, except for indicating that oxygen was available on descent meaning there wasn't excessive use during ascent. It is worth noting that most fatalities on Everest happen during the descent.
The next features that were slightly more central in successful summits were previous experience about 8000m (for reference Everest is at 8849m), followed by use of O2 while ascending. Surprisingly, summit centrality for sex (indicating male) was relatively low compared to no-summit centrality indicating that being male had low importance in the chances of success at summit. As expected a major difference in Lastly, the largest differences in summit vs no summit were from identity (sherpa were much more likely to succeed), and age ($<40$ year olds were much more likely to succeed), as expected and also seen in previous studies \cite{huey2020}. 
Fig. \ref{fig:centrality} shows the intra-expedition graphs averaged over groups of successful (b) vs unsuccessful (c) climbers. Note that these graphs are projected from a climber-feature bipartite network, and hence encode the true distribution of features.

\section{Generating a multiscale network}
\label{sec:correlation}
\begin{figure}[htbp!]
	\centering
	\includegraphics[width=0.9\linewidth]{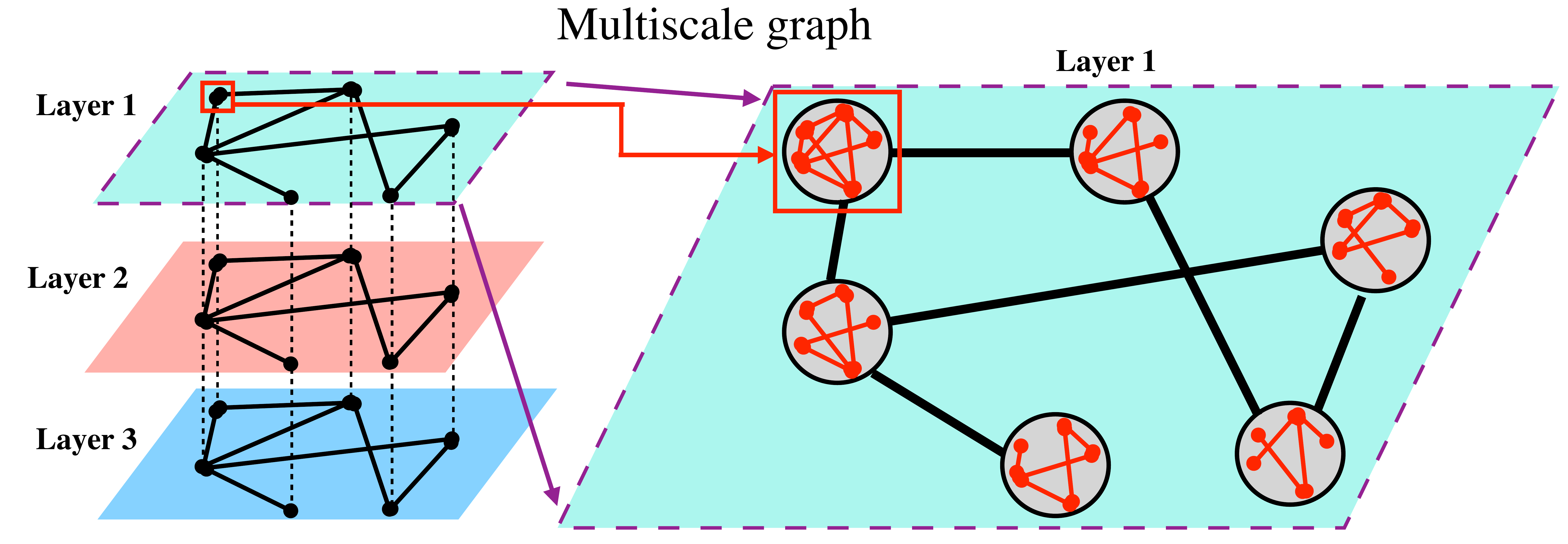}
	\caption{An illustrative example of a multiscale graph.}
	\label{fig:multiscale_illus}
\end{figure}

Graphs are a natural way to model relationships or comparisons between expeditions. Multilayer or multiplex networks are commonly used to model relationships of multiple types, where each layer corresponds to a specific `type'. While the intra-expedition network provides insight into the features that climbers in the expedition shared, expedition-wide factors also determine the success of an expedition. The expedition-wide factors considered here are: (1) number of days to summit from base camp, (2) number of high points/camps, (3) expedition size (including hired personnel), (4) ratio of number of paying climbers to number of hired personnel. In order to compare different expeditions, one must consider both the intra-expedition feature similarity as well as the 4 expedition-wide factors. We model each of these as a layer in a multilayer expedition similarity graph, resulting in 5 total layers.
Thus the multiplex graph is also a multiscale graph $E$ where all expeditions are represented as the nodes of the multiplex graph, however, within the first layer, the nodes themselves are the intra-layer feature graphs, lending to it a hierarchical or multiscale structure. An illustration of this is shown in Fig. \ref{fig:multiscale_illus}.

Now, in practice the multiscale multiplex graph $E$ is generated as follows: 
For the four layers encoding expedition-wide factors, expedition $i$ is connected to expedition $j$ in layer $l$ if both expedition $i$ and expedition $j$ have a value of layer $l$ greater than the mean of the layer $\mu^l$. In other words, 
\begin{align}
	A_{ij}^l &= 1 \textup{ if } v_{i}^l, v_j^l > \mu^l, \\
	&= 0 \textup{ otherwise.}
\end{align}
where $v_i^l$ is the value of factor $l$ taken by expedition $i$. 
For the intra-expedition feature layer, the connection between two expeditions is given by the Graph Edit distance \cite{gao2010} between their intra-expedition graphs. This is then normalized to have a max value of 1.

\section{Determining layer importance through correlation with success}

Different factors encoded in the layers may varying importance in determining the success of an expedition.
Each expedition has a success rate that is computed from the fraction of climbers that succeeded at summiting. The importance of each layer can be inferred from the correlation between the its values and the success rates. However, for the intra-expedition layer, this involves computing the correlations between graphs and success rate.

\begin{figure}[htbp!]
	\centering
	\includegraphics[width=0.75\linewidth]{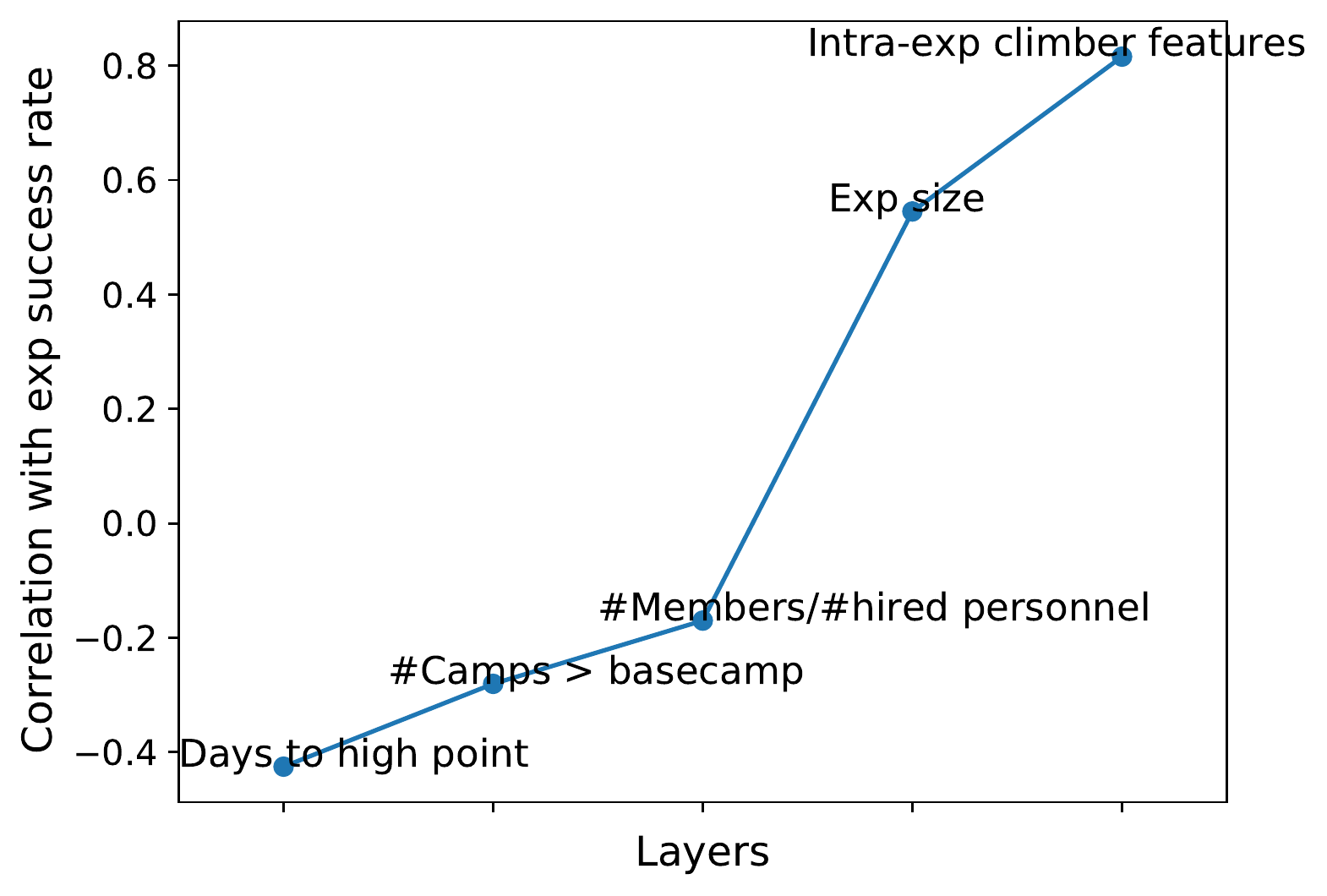}
	\caption{Pearson's correlation coefficient between layer (factor) values and expedition success rate. The exact values across x-axis layers are $-0.45, -0.36, -0.12, 0.57, 0.84$. The corresponding $p$-values are $5.5 \times 10^{-10}, 1.15 \times 10^{-6}, 0.1, 5.7 \times 10^{-16}, 8.9 \times 10^{-47}$.}
	\label{fig:correlation}
\end{figure}

An important problem in network science involves devising measures to compare graphs to scalars. Here, this problem manifests in terms of computing correlation between the intra-expedition layer (comprising of graphs) and their corresponding scalar success rate. Since each factor in the graph is independent, we perform linear regression on the unique entries of the adjacency matrix of the graph to obtain a linear fit that best map the intra-expedition graphs to the success rates. The corresponding coefficients are denoted by $\vec{c}$. Note that linear regression generates a single set of coefficients that are best map every graph to its scalar output. In principle one can use higher order methods, or neural networks to learn this mapping, however, since the features of the graph are expected to be linearly independent, considering higher order terms is unnecessary.
Now, in order to calculate the correlation between the graphs and scalars, one can project the graphs as follows: the intra-expedition graph of the $j^{th}$ expedition whose adjacency matrix is $A^{j}$ is represented by the scalar $\eta_j$ given by $\eta_j = A^{j} \cdot \vec{c}$. One can then identify the importance of the intra-expedition layer through the correlation coefficient between $\vec{\eta}$ and the success rates. 
Fig. \ref{fig:correlation} shows the Pearson's correlation coefficient between the layers and the success rate. A higher correlation implies higher influence of the layer in determining success. Surprisingly, the factor with the least importance (smallest absolute value of correlation) was the ratio of number of paying members to number of hired personnel on the team. Both number of camps above basecamp and days to summit/high point had a negative correlation with success, as one might expect, with the latter having a larger effect. Also surprisingly, the \textit{expedition size} is found to be relatively important in determining success (with a correlation coefficient of $>$ 0.5). Lastly, the most important factor was the intra-expedition feature graph layer which is strongly correlated with success, indicating that non-linear effects and outliers to the regression fit are relatively few. The p-value for the layers are extremely low indicating that the correlation is statistically significant except for the number of members to hired personnel.

\section{Community detection to identify patterns of success}

\begin{figure}[htbp!]
	\centering
	\includegraphics[width=0.48\linewidth]{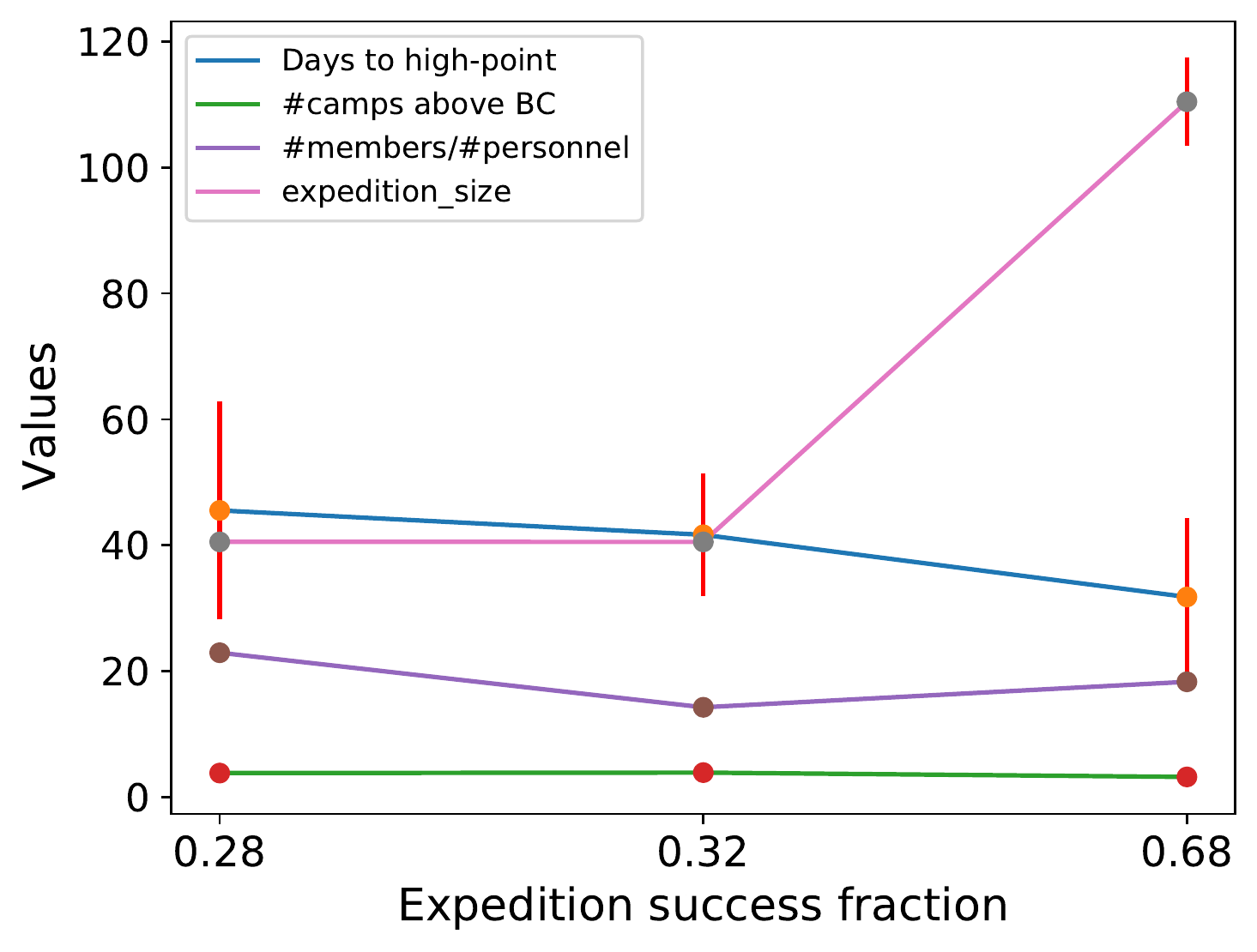}
	\includegraphics[width=0.48\linewidth]{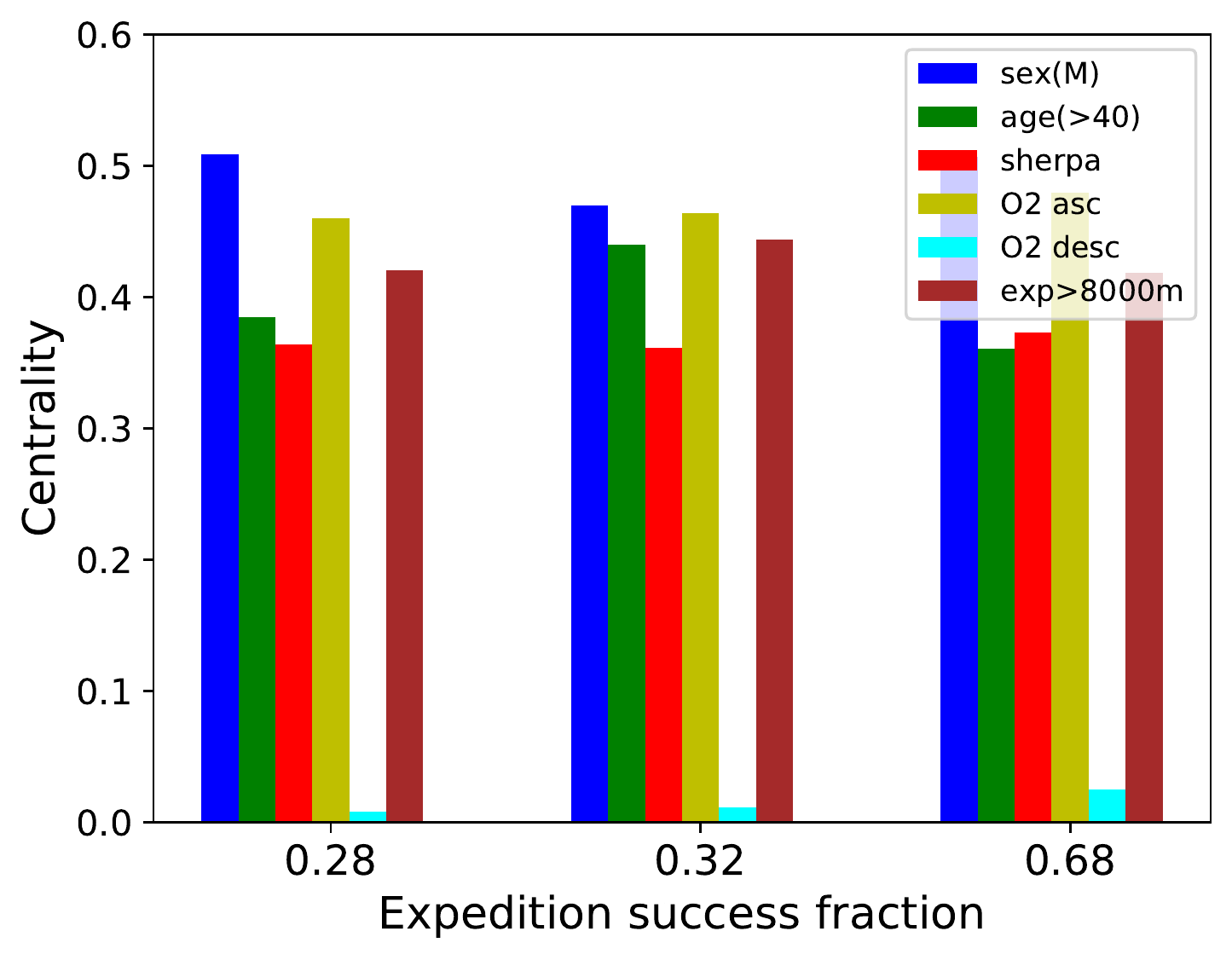}
	\caption{(Left) The average values of the various expedition-wide factors represented in layers shown for the three communities. (right) Centralities of the intra-expedition graph features for the three communities. The three communities are represented by their success fraction across the x axis.}
	\label{fig:communities}
\end{figure}

Success at high peaks is a combination of several features. By developing an expedition similarity graph, one can identify communities of expeditions that have similar factors and features. One may wonder if the data would naturally cluster into communities with different success rates, which can be associated to differences in the combination of factors. The layers of the multiscale multiplex graph $E$ are aggregated to generate an expedition similarity graph $S$ given by
$S=\sum_l E^l/|l|$, where $l$ is the total number of layers (5 in this case).
Here, each layer has the same weight, but one may choose to weight them other ways, for instance weighted by layer importance. Louvain community detection \cite{blondel2008} is then applied to $S$ and identifies three communities. Note that the number of communities is not pre-determined but selected by the algorithm maximize modularity \cite{newman2006}. Fig. \ref{fig:communities} (left) shows the differences in expedition-wide factors in the three communities (ordered by average success rate on the x-axis). As seen from the figure, the three emergent communities naturally bifurcated to reveal different success rates (the first and the second community had similar success rates at 0.28 and 0.32 whereas the third was significantly higher at 0.68). Firstly, the most dominant difference is found to be expedition size which is significantly higher in the third `successful' community (with the highest success rate at 0.68), indicating that larger groups allow a larger fraction of climbers to succeed. One may hypothesize that this is because the experienced climbers do not have to shoulder the responsibility of the less experienced climbers, which may slow them down.
Additionally, All communities were largely similar in the number of camps above base camp. However, the ratio of number of members to personnel was the relatively higher in both the community with the highest success rate, as well as the lowest success rate, indicating that it isn't a determining factor, supported by results from section \ref{sec:correlation}. Lastly, as the success rate of the communities increase, their days to summit decrease, which is also expected. At high altitudes, the body goes into shock from prolonged exposure, so one might expect a faster expedition to face less challenges in this regard, and hence be more successful.

Fig. \ref{fig:communities} (right) plots the centralities of the intra-expedition features for the three communities (ordered by their success rate on the x-axis). The successful community, were the youngest and had the highest centrality for oxygen use both while ascending and while descending. Despite having slightly less average experience than the first two groups, \textit{age} and \textit{oxygen} use were the leading indicative features for success, which is in agreement with results from \ref{sec:centrality}. 
The main differences between the first and second `low-success' communities are that the first community had relatively low experience $>$8000m, whereas the second is a relatively older, but more experienced population.

\section{Discussion and Future work}

This work presents the first network-based analysis of mountaineering, studying the intra-expedition and expedition-wide factors that contribute to success. First, it considers a climber-centric perspective and shows that the chances of summit failure (due to fatigue, logistical failure etc.) drastically reduce when climbing with repeat partners, especially for more experienced climbers. Then, it studies the importance of intra-expedition features by projecting a bipartite climber-feature network to show that the largest different in centralities amongst successful and unsuccessful groups is found in the `age' node, indicating that it's the strongest driver of success. Further, it introduces a multiscale multiplex network to model similarities between expeditions, where one or more layers may be multiscale whereas others are not. Such a multiscale approach can model a variety of systems, and the tools used here to navigate simultaneous modeling of different types of layers and project networks to a scalar space through regression are applicable in a variety of scenarios. Lastly, community detection on the expedition-similarity reveals three distinct communities where a difference in success rates naturally emerges amongst the communities. The dominant characteristics that support a successful outcome for a large fraction of the expedition are high expedition size, low age, and oxygen use.
Future work may include study of additional factors, analysis of death factors, and a multilayer or multiscale approaches to modularity optimization and community detection. Code can be found at \textit{https://github.com/chimeraki/Multiscale\textunderscore network\textunderscore mountaineering}.


\bibliographystyle{abbrv}    
\bibliography{himalaya}   

\begin{thebibliography}{10}

\bibitem{betzel2017}
R.~F. Betzel and D.~S. Bassett.
\newblock Multi-scale brain networks.
\newblock {\em Neuroimage}, 160:73--83, 2017.

\bibitem{blondel2008}
V.~D. Blondel, J.-L. Guillaume, R.~Lambiotte, and E.~Lefebvre.
\newblock Fast unfolding of communities in large networks.
\newblock {\em Journal of statistical mechanics: theory and experiment},
  2008(10):P10008, 2008.

\bibitem{buldu2018}
J.~M. Buld{\'u}, J.~Busquets, J.~H. Mart{\'\i}nez, J.~L. Herrera-Diestra,
  I.~Echegoyen, J.~Galeano, and J.~Luque.
\newblock Using network science to analyse football passing networks: Dynamics,
  space, time, and the multilayer nature of the game.
\newblock {\em Frontiers in psychology}, 9:1900, 2018.

\bibitem{ewert1985}
A.~Ewert.
\newblock Why people climb: The relationship of participant motives and
  experience level to mountaineering.
\newblock {\em Journal of Leisure Research}, 17(3).

\bibitem{gao2010}
X.~Gao, B.~Xiao, D.~Tao, and X.~Li.
\newblock A survey of graph edit distance.
\newblock {\em Pattern Analysis and applications}, 13(1):113--129, 2010.

\bibitem{hammoud2020}
Z.~Hammoud and F.~Kramer.
\newblock Multilayer networks: aspects, implementations, and application in
  biomedicine.
\newblock {\em Big Data Analytics}, 5(1):1--18, 2020.

\bibitem{helms1984}
M.~Helms.
\newblock Factors affecting evaluations of risks and hazards in mountaineering.
\newblock {\em Journal of Experiential Education}, 7(3):22--24, 1984.

\bibitem{huang2021}
X.~Huang, D.~Chen, T.~Ren, and D.~Wang.
\newblock A survey of community detection methods in multilayer networks.
\newblock {\em Data Mining and Knowledge Discovery}, 35(1):1--45, 2021.

\bibitem{huey2020}
R.~B. Huey, C.~Carroll, R.~Salisbury, and J.-L. Wang.
\newblock Mountaineers on mount everest: Effects of age, sex, experience, and
  crowding on rates of success and death.
\newblock {\em Plos one}, 15(8):e0236919, 2020.

\bibitem{huey2001}
R.~B. Huey and X.~Eguskitza.
\newblock Limits to human performance: elevated risks on high mountains.
\newblock {\em Journal of Experimental Biology}, 204(18):3115--3119, 2001.

\bibitem{huey2007}
R.~B. Huey, R.~Salisbury, J.-L. Wang, and M.~Mao.
\newblock Effects of age and gender on success and death of mountaineers on
  mount everest.
\newblock {\em Biology Letters}, 3(5):498--500, 2007.

\bibitem{kivela2014}
M.~Kivel{\"a}, A.~Arenas, M.~Barthelemy, J.~P. Gleeson, Y.~Moreno, and M.~A.
  Porter.
\newblock Multilayer networks.
\newblock {\em Journal of complex networks}, 2(3):203--271, 2014.

\bibitem{krishnagopal2021}
S.~Krishnagopal.
\newblock Multi-layer trajectory clustering: A network algorithm for disease
  subtyping.
\newblock {\em Biomedical Physics \& Engineering Express}, 6(6):065003, 2020.

\bibitem{krishnagopal2020}
S.~Krishnagopal, R.~v. Coelln, L.~M. Shulman, and M.~Girvan.
\newblock Identifying and predicting parkinson’s disease subtypes through
  trajectory clustering via bipartite networks.
\newblock {\em PloS one}, 15(6):e0233296, 2020.

\bibitem{krishnagopal2017}
S.~Krishnagopal, J.~Lehnert, W.~Poel, A.~Zakharova, and E.~Sch{\"o}ll.
\newblock Synchronization patterns: From network motifs to hierarchical
  networks.
\newblock {\em Philosophical Transactions of the Royal Society A: Mathematical,
  Physical and Engineering Sciences}, 375(2088):20160216, 2017.

\bibitem{larremore2014}
D.~B. Larremore, A.~Clauset, and A.~Z. Jacobs.
\newblock Efficiently inferring community structure in bipartite networks.
\newblock {\em Physical Review E}, 90(1):012805, 2014.

\bibitem{lee2015}
K.-M. Lee, B.~Min, and K.-I. Goh.
\newblock Towards real-world complexity: an introduction to multiplex networks.
\newblock {\em The European Physical Journal B}, 88(2):1--20, 2015.

\bibitem{lenormand2018}
M.~Lenormand, S.~Luque, J.~Langemeyer, P.~Tenerelli, G.~Zulian, I.~Aalders,
  S.~Chivulescu, P.~Clemente, J.~Dick, J.~Van~Dijk, et~al.
\newblock Multiscale socio-ecological networks in the age of information.
\newblock {\em PloS one}, 13(11):e0206672, 2018.

\bibitem{lusher2010}
D.~Lusher, G.~Robins, and P.~Kremer.
\newblock The application of social network analysis to team sports.
\newblock {\em Measurement in physical education and exercise science},
  14(4):211--224, 2010.

\bibitem{mo2019}
H.~Mo and Y.~Deng.
\newblock Identifying node importance based on evidence theory in complex
  networks.
\newblock {\em Physica A: Statistical Mechanics and its Applications},
  529:121538, 2019.

\bibitem{newman2006}
M.~E. Newman.
\newblock Modularity and community structure in networks.
\newblock {\em Proceedings of the national academy of sciences},
  103(23):8577--8582, 2006.

\bibitem{nyaupane2015}
G.~Nyaupane, G.~Musa, J.~Higham, and A.~Thompson-Carr.
\newblock Mountaineering on mt everest: evolution, economy, ecology and ethics.
\newblock {\em Mountaineering tourism, Routledge, New York, NY}, page 265,
  2015.

\bibitem{pereira2019}
E.~J. d. A.~L. Pereira, P.~J.~S. Ferreira, M.~F. da~Silva, J.~G.~V. Miranda,
  and H.~Pereira.
\newblock Multiscale network for 20 stock markets using dcca.
\newblock {\em Physica A: Statistical Mechanics and its Applications},
  529:121542, 2019.

\bibitem{pomfret2006}
G.~Pomfret.
\newblock Mountaineering adventure tourists: a conceptual framework for
  research.
\newblock {\em Tourism management}, 27(1):113--123, 2006.

\bibitem{saito2016}
K.~Saito, M.~Kimura, K.~Ohara, and H.~Motoda.
\newblock Super mediator--a new centrality measure of node importance for
  information diffusion over social network.
\newblock {\em Information Sciences}, 329:985--1000, 2016.

\bibitem{data04}
R.~Salisbury.
\newblock The himalayan database: the expedition archives of elizabeth hawley.
\newblock {\em Golden: American Alpine Club}, 2004.

\bibitem{sarkar2013}
S.~Sarkar, J.~A. Henderson, and P.~A. Robinson.
\newblock Spectral characterization of hierarchical network modularity and
  limits of modularity detection.
\newblock {\em PloS one}, 8(1):e54383, 2013.

\bibitem{schussman1990}
L.~Schussman, L.~Lutz, R.~Shaw, and C.~Bohnn.
\newblock The epidemiology of mountaineering and rock climbing accidents.
\newblock {\em Journal of Wilderness Medicine}, 1(4):235--248, 1990.

\bibitem{sola2013}
L.~Sol{\'a}, M.~Romance, R.~Criado, J.~Flores, A.~Garc{\'\i}a~del Amo, and
  S.~Boccaletti.
\newblock Eigenvector centrality of nodes in multiplex networks.
\newblock {\em Chaos: An Interdisciplinary Journal of Nonlinear Science},
  23(3):033131, 2013.

\bibitem{steinhaeuser2011}
K.~Steinhaeuser, N.~V. Chawla, and A.~R. Ganguly.
\newblock Complex networks as a unified framework for descriptive analysis and
  predictive modeling in climate science.
\newblock {\em Statistical Analysis and Data Mining: The ASA Data Science
  Journal}, 4(5):497--511, 2011.

\bibitem{szymczak2021}
R.~K. Szymczak, M.~Marosz, T.~Grzywacz, M.~Sawicka, and M.~Naczyk.
\newblock Death zone weather extremes mountaineers have experienced in
  successful ascents.
\newblock {\em Frontiers in Physiology}, 12:998, 2021.

\bibitem{tempest2007}
S.~Tempest, K.~Starkey, and C.~Ennew.
\newblock In the death zone: A study of limits in the 1996 mount everest
  disaster.
\newblock {\em Human Relations}, 60(7):1039--1064, 2007.

\bibitem{tougne2008}
J.~Tougne, B.~Paty, D.~Meynard, J.-M. Martin, T.~Letellier, and E.~Rosnet.
\newblock Group problem solving and anxiety during a simulated mountaineering
  ascent.
\newblock {\em Environment and Behavior}, 40(1):3--23, 2008.

\bibitem{vaiana2020}
M.~Vaiana and S.~F. Muldoon.
\newblock Multilayer brain networks.
\newblock {\em Journal of Nonlinear Science}, 30(5):2147--2169, 2020.

\bibitem{weinbruch2013}
S.~Weinbruch and K.-C. Nordby.
\newblock Fatalities in high altitude mountaineering: A review of quantitative
  risk estimates.
\newblock {\em High altitude medicine \& biology}, 14(4):346--359, 2013.

\bibitem{westhoff2012}
J.~L. Westhoff, T.~D. Koepsell, and C.~T. Littell.
\newblock Effects of experience and commercialisation on survival in himalayan
  mountaineering: retrospective cohort study.
\newblock {\em bmj}, 344, 2012.

\bibitem{disease2019}
A.~P. y~Piontti, N.~Perra, L.~Rossi, N.~Samay, and A.~Vespignani.
\newblock {\em Charting the next pandemic: modeling infectious disease
  spreading in the data science age}.
\newblock Springer, 2019.

\end{thebibliography}

\end{document}